\documentclass[aps,prl,preprint,groupedaddress,showpacs]{revtex4-1}
\usepackage[latin9]{inputenc}
\usepackage{verbatim}
\usepackage{esint}

\makeatletter
 {%
 }

\makeatletter

\begin{document}


\title{Development of Eulerian Theory of Turbulence within Kraichnan's Direct
Interaction Approximation Framework}



\author{R. V. R. Pandya}
\email{rvrptur\{AT\}yahoo.com, raja.pandya\{AT\}upr.edu}

\affiliation{Department of Mechanical Engineering, University of Puerto Rico at
Mayaguez, Mayaguez, Puerto Rico, PR 00681, USA}

\maketitle
{\bf ABSTRACT}

Within the framework of Kraichnan's Direct Interaction Approximation
(DIA), we propose an Eulerian turbulence theory providing a closed
set of equations for two-time and single-time velocity correlations,
and second order correlations of infinitesimal response tensor $\hat{G}_{in}({\bf k};t,t')$. The proposed theory, namely variant of DIA (VDIA), is consistent with Kolmogorov's energy spectrum. The VDIA is further modified to make it compatible with 
random Galilean transformation rules. The closed set of equations does not contain equation for ensemble averaged response
tensor $G_{in}({\bf k};t,t')=\langle\hat{G}_{in}({\bf k};t,t')\rangle$. The present theory can also be seen as a new remormalized perturbation theory having different method for renormalization.


PACS: 47.27.Ak, 47.27.eb, 47.27.ef, 47.27.Gs




\section{Introduction}

Kraichnan \cite{Kraichnan58,Kraichnan59} proposed Direct Interaction
Approximation (DIA) as a pioneer renormalized perturbation theory
(RPT) to solve turbulence closure problem. The DIA has been of central importance to Kraichnan's work \cite{KH78} and many other contibutions in subject turbulence. After DIA, various turbulence
theories \cite{Wyld61,Edwards64,Kraichnan65a,Lee65,Herring65,Herring66,EM69,Phythian69,
BS70,Orszag70,Nakano72,Nakano88,MSR73,FLB74,FNS77,KH78,McComb78,DM79,HL79,Kaneda81,
AVP83,Qian83,Lvov91,BKO93,MW95,LP95,LP95b,Eijnden97,KG97,Pandya04a,Pandya14}
have been proposed by researchers and many have been reviewed from
time to time \cite{Leslie73,McComb90,McComb95,Lvov91,Lesieur97,Krommes97,Yoshizawa98,AAV99}.

Kraichnan introduced infinitesimal response tensor in DIA which plays
a central role in achieving closure solution. The DIA solution for
homogeneous, isotropic turbulence consists of closed set of three equations for
ensemble averaged response function, single-time and two-time velocity
correlations. The DIA theory is energetically consistent \cite{Kraichnan61}
and has model amplitude equation representation \cite{Phythian69,Kraichnan70a}.
It accurately predicts decay of isotropic turbulence at moderate Reynolds number.
The DIA and its related theories have been applied to real turbulent flows \cite{Kraichnan64b,Kraichnan64c,Leslie73,Frederiksen99,OF2004,OF2010}. 

However, the DIA failed to reproduce Kolmogorov's $k^{-5/3}$ law
for energy spectrum \cite{Kolmogorov41,Batchelor59,Frisch95} which
exists at high Reynolds number. The failure is due to divergence of
the response function equation \cite{Leslie73} when Kolmogorov's
spectrum is used in it. Such divergence is not present in DIA equations
for single-time and two-time velocity correlations. Yet another reason
for the failure was suggested by Kraichnan \cite{Kraichnan64}. He
argued that Eulerian framework of DIA is unsuited as Eulerian moments
can not properly separate and account for two different mechanisms:
1) sweeping of small-scale structures by large-scales and 2) intrinsic,
internal distortion of small scales. In particular, DIA is inaccurate
in capturing the sweeping effect \cite{Kraichnan64}. Later, Kraichnan
\cite{Kraichnan65a} provided mathematical interpretation of the sweeping
effect which is known as invariance under random Galilean transformation
(RGT). 

In a nutshell, we see two main deficiencies in DIA theory and which are:
1) the existence of divergence in response function Eq. (\ref{eq:21}) when
Kolmogorov spectrum is substituted in it, and
2) DIA's incompatibility with random Galilean transformation (RGT).
Kraichnan tackled these deficiencies by implementing DIA in generalized
Lagrangian framework which resulted in a theory known as Lagrangian
history direct interaction (LHDI) \cite{Kraichnan65a}. The LHDI was
further improved for quantitative prediction, resulting in strain-based
Lagrangian-history direct interaction (SBLHDI) theory \cite{KH78,HK79}.
The LHDI is compatible with transformation rules as provided
by RGT \cite{Kraichnan65a} and is successfully applied to
multiphase turbulence \cite{Reeks91,Reeks92}. There are other Lagrangian \cite{Kaneda81,KG97} and
semi-Lagrangian \cite{BL87,Lvov91} theories which are compatible
with Kolmogorov energy spectrum. In particular, semi-Lagrangian theory
by L'vov and his coworkers \cite{BL87,LP95,LP95b} is most successful
in removing the sweeping effect completely and predicting anomalous
scaling. Their theory is in agreement with random Galilean transformation
as it removes the sweeping effect in instantaneous equation for semi-Lagrangian
velocity field.

In the present theory, we seek to obtain a closure solution in Eulerian framework for properly quantifying internal distortion and sweeping mechanisms simultaneously. This would allow the theory to be extended to general flow situations where sweeping effect is equally important. Over a period of time, I have been studying the work of Kraichnan and I strongly believe that a successful theory of turbulence is hidden within his work. To turn this belief into reality, as a first step, here we propose an Eulerian theory to tackle two main deficiencies of DIA theory of central importance.  

Now we present basic equations and set of closure equations of DIA. Then we present the new Eulerian theory, namely variant of DIA (VDIA), its modifications which make it compatible with RGT and concluding remarks.

\section{Basic equations and DIA closure equations}

\subsection{Navier-Stokes equation and closure problem}

Consider a homogeneous, isotropic, incompressible fluid turbulence
inside a cubic box of side $D$ in a reference frame $S$ which is
stationary in the laboratory. The Eulerian velocity field $u_{i}({\bf x},t)$
in physical space-time (${\bf x},t$) can be expressed in terms of
Fourier modes of the velocity field $u_{i}({\bf k},t)$ by using the
Fourier transform

\begin{equation}
u_{i}({\bf x},t)=\sum_{{\bf k}}u_{i}({\bf k},t)\exp(\iota{\bf k.x}),
\end{equation}
where $\iota=\sqrt{-1}$ and summation is taken over all permitted
wave-vectors ${\bf k}$ and subscripts take the value 1, 2 or 3. The
Fourier transform of Navier-Stokes equations governing the dynamics
of $u_{i}({\bf x},t)$ can be written as \cite{Kraichnan59,Leslie73}
\begin{equation}
\left(\frac{\partial}{\partial t}+\nu k^{2}\right)u_{i}({\bf k},t)=M_{ijm}({\bf k})\sum_{{\bf {\bf p+{\bf q=}}k}}u_{j}({\bf p},t)u_{m}({\bf {\bf q}},t),\label{eq:2}
\end{equation}
and the continuity equation is 
\begin{equation}
k_{j}u_{j}({\bf k},t)=0.
\end{equation}
Here $\nu$ is kinematic viscosity of the fluid and 
\begin{equation}
M_{ijm}({\bf k})=-\frac{\iota}{2}\left[k_{m}P_{ij}({\bf k})+k_{j}P_{im}({\bf k})\right]
\end{equation}
and by using Kronecker delta $\delta_{ij}$, the projector $P_{ij}({\bf k})$
can be written as 
\begin{equation}
P_{ij}({\bf k})=\delta_{ij}-k_{i}k_{j}|{\bf k}|^{-2}.
\end{equation}

Denoting the operation of taking an ensemble average by $\langle\,\,\rangle$,
equations governing the evolution of two-time velocity correlation
\begin{equation}
Q_{in}({\bf k},{\bf -k};t,t')=\langle u_{i}({\bf k},t)u_{n}({\bf -k},t')\rangle
\end{equation}
and single-time velocity correlation 
\begin{equation}
Q_{in}({\bf k},{\bf -k};t,t)=\langle u_{i}({\bf k},t)u_{n}({\bf -k},t)\rangle
\end{equation}
can be obtained from Eq. (\ref{eq:2}) and written as 
\begin{equation}
\left(\frac{\partial}{\partial t}+\nu k^{2}\right)Q_{in}({\bf k},{\bf -k};t,t')=M_{ijm}({\bf k})\sum_{{\bf {\bf p+{\bf q=}}k}}\langle u_{j}({\bf p},t)u_{m}({\bf q},t)u_{n}({\bf -k},t')\rangle,\label{eq:8}
\end{equation}
and 
\begin{eqnarray}
\left(\frac{\partial}{\partial t}+2\nu k^{2}\right)Q_{in}({\bf k},{\bf -k};t,t)=M_{ijm}({\bf k})\sum_{{\bf {\bf p+{\bf q=}}k}}\langle u_{j}({\bf p},t)u_{m}({\bf q},t)u_{n}(-{\bf k},t)\rangle\nonumber \\
+M_{njm}({\bf -k})\sum_{{\bf {\bf p+{\bf q=-}}k}}\langle u_{j}({\bf p},t)u_{m}({\bf q},t)u_{i}({\bf k},t)\rangle.\label{eq:9}
\end{eqnarray}
These Eqs. (\ref{eq:8}) and (\ref{eq:9}) pose well known closure
problem of turbulence due to the presence of unknown third-order velocity
correlations.

For homogeneous, isotropic turbulence these correlations can be written
as 
\begin{equation}
\left(\frac{D}{2\pi}\right)^{3}Q_{in}({\bf k},{\bf -{\bf k}};t,t')=\frac{1}{2}P_{in}({\bf k})U(k;t,t'),\label{eq:10}
\end{equation}
\begin{equation}
\left(\frac{D}{2\pi}\right)^{3}Q_{in}({\bf k},{\bf -{\bf k}};t,t)=\frac{1}{2}P_{in}({\bf k})U(k;t,t),\label{eq:11}
\end{equation}
 and equations for $U(k;t,t')$ and $U(k;t,t)$ can be obtained from
Eqs. (\ref{eq:8}) and (\ref{eq:9}), respectively. These equations
may be written as

\begin{equation}
\left[\frac{\partial}{\partial t}+\nu k^{2}\right]U(k;t,t')=S(k;t,t'),\label{eq:12}
\end{equation}
\begin{equation}
\left[\frac{\partial}{\partial t}+2\nu k^{2}\right]U(k;t,t)=2S(k;t,t),\label{eq:13}
\end{equation}
where 
\begin{equation}
S(k;t,t')=\left(\frac{D}{2\pi}\right)^{3}M_{ijm}({\bf k})\sum_{{\bf {\bf p+{\bf q=}}k}}\langle u_{j}({\bf p},t)u_{m}({\bf q},t)u_{i}({\bf -k},t')\rangle\label{eq:14}
\end{equation}

\subsection{DIA equations}

Kraichnan introduced infinitesimal response
tensor $\hat{G}_{in}({\bf k};t,t')$ for Eq. (\ref{eq:2}) in DIA to solve closure problem \cite{Kraichnan58,Kraichnan59}. The governing equation for $\hat{G}_{in}({\bf k};t,t')$
can be written as \cite{Leslie73,McComb90} 
\begin{equation}
\left(\frac{\partial}{\partial t}+\nu k^{2}\right)\hat{G}_{in}({\bf k};t,t')-2M_{ijm}({\bf k})\sum_{{\bf {\bf p+{\bf q=}}k}}u_{j}({\bf p},t)\hat{G}_{mn}({\bf {\bf q}};t,t')=P_{in}({\bf k})\delta(t-t')\label{eq:15}
\end{equation}
For homogeneous, isotropic turbulence, ensemble averaged response
tensor can be written as 
\begin{equation}
G_{in}({\bf k};t,t')=\langle\hat{G}_{in}({\bf k};t,t')\rangle=P_{in}({\bf k})G(k;t,t').
\end{equation}
The equation for $G(k;t,t')$ for $t>t'$ may be obtained from Eq.
(\ref{eq:15}) and written as
\begin{equation}
\left[\frac{\partial}{\partial t}+\nu k^{2}\right]G(k;t,t')=C(k;t,t'),\label{eq:17}
\end{equation}
where 
\begin{equation}
C(k;t,t')=M_{ijm}({\bf k})\sum_{{\bf {\bf p+{\bf q=}}k}}\langle u_{j}({\bf p},t)\hat{G}_{mi}({\bf {\bf q}};t,t')\rangle.\label{eq:18}
\end{equation}

The DIA theory yields closed set of equations for $Q_{in}({\bf k},{\bf -k};t,t')$,
$Q_{in}({\bf k},{\bf -k};t,t)$, $G_{in}({\bf k};t,t')$ and provides
closure solution for $S(k;t,t')$ and $C(k;t,t')$. For homogeneous, isotropic
turbulence, DIA equations for exact Eqs. (\ref{eq:12})-(\ref{eq:14}),
(\ref{eq:17}), (\ref{eq:18}) may be written as \cite{Kraichnan59,Kraichnan64a,Leslie73}
\begin{equation}
\left[\frac{\partial}{\partial t}+\nu k^{2}\right]U(k;t,t')=S^{DIA}(k;t,t'),\label{eq:19}
\end{equation}
\begin{equation}
\left[\frac{\partial}{\partial t}+2\nu k^{2}\right]U(k;t,t)=2S^{DIA}(k;t,t),\label{eq:20}
\end{equation}
\begin{eqnarray}
\left[\frac{\partial}{\partial t}+\nu k^{2}\right]G(k;t,t') & =C^{DIA}(k;t,t') & =-\pi k\int\int_{\Delta}\mathrm{d}p\mathrm{d}qpqb(k,p,q)\Bigl[\nonumber \\
 &  & \int_{t'}^{t}G(k;s,t')G(p;t,s)U(q;t,s)\mathrm{d}s\Bigr],\:\:\forall t>t'.\label{eq:21}
\end{eqnarray}
Here 
\begin{eqnarray}
S^{DIA}(k;t,t') & = & \pi k\int\int_{\Delta}\mathrm{d}p\mathrm{d}qpq\Bigl[\int_{0}^{t'}a(k,p,q)G(k;t',s)U(p;t,s)U(q;t,s)\mathrm{d}s\nonumber \\
 &  & -\int_{0}^{t}b(k,p,q)U(k;t',s)G(p;t,s)U(q;t,s)\mathrm{d}s\Bigr],\label{eq:22}
\end{eqnarray}
\begin{equation}
a(k,p,q)=\frac{1}{2}(1-xyz-2y^{2}z^{2}),
\end{equation}
\begin{equation}
b(k,p,q)=(p/k)(xy+z^{3}),
\end{equation}
$x$, $y$ and $z$ are cosines of interior angles opposite
of sides $k$, $p$ and $q$ of a triangle, respectively. 

The rhs of Eq. (\ref{eq:21}) diverges for Kolmogorov's energy spectrum
\cite{Leslie73}. The Eqs. (\ref{eq:19}), (\ref{eq:20}) along with
Eq. (\ref{eq:22}) are consistent with Kolmogorov's energy spectrum
and the divergence in these equations cancels out between two terms
present on the rhs of Eq. (\ref{eq:22}) \cite{Leslie73}. According
to S. F. Edwards, in his private communication to Leslie \cite{Leslie73},
Eq. (\ref{eq:21}) would have converged if it had an additional term
similar to first term on the rhs of Eq. (\ref{eq:22}). In the present theory, such an additional term exists in equation for  a function $K(k;t,t')$ (please see Eqs. (\ref{eq:36}), (\ref{eq:37})) which replaces DIA Eq. (\ref{eq:21}).

\section{Present Eulerian Turbulence Theory: Variant of DIA}

The present theory, namely, variant of DIA (VDIA) utilizes second
order correlations of infinitesimal response tensor $\hat{G}_{in}({\bf k};t,t')$
which are defined as
\begin{equation}
G_{inab}^{G}({\bf k};t,t'|-{\bf k};t,t')=\langle\hat{G}_{in}({\bf k};t,t')\hat{G}_{ab}({\bf -k};t,t')\rangle
\end{equation}
 and 
\begin{equation}
G_{inab}^{G}({\bf k};t,t'|-{\bf k};t'',t')=\langle\hat{G}_{in}({\bf k};t,t')\hat{G}_{ab}({\bf -k};t'',t')\rangle.
\end{equation}
For homogeneous, isotropic turbulence, we consider 
\begin{equation}
G_{inab}^{G}({\bf k};t,t'|-{\bf k};t'',t')=P_{in}({\bf k})P_{ab}(-{\bf k})G^{G}(k;t,t'|t'',t')
\end{equation}
and 
\begin{equation}
G^{G}(k;t,t'|t,t')=\left[K(k;t,t')\right]^{2}.
\end{equation}
Also, we define response correlation function $R^{G}(k;t,t'|t'',t')$
for $t>t'$ and $t''>t'$ as 
\begin{equation}
R^{G}(k;t,t'|t'',t')=\frac{G^{G}(k;t,t'|t'',t')}{K(k;t,t')K(k;t'',t')}.\label{eq:29-1}
\end{equation}
It should be noted that $R^{G}(k;t,t'|t,t')=1.$ 

The exact Eqs. for $G^{G}(k;t,t'|t,t')$ and $G^{G}(k;t,t'|t'',t')$
for $t>t',\, t''>t'$ may be obtained from Eq. (\ref{eq:15}) and
written as
\begin{equation}
\left(\frac{\partial}{\partial t}+2\nu k^{2}\right)\left[K(k;t,t')\right]^{2}=2L(k;t,t'|t,t'),\label{eq:30-1}
\end{equation}
\begin{equation}
\left(\frac{\partial}{\partial t}+\nu k^{2}\right)G^{G}(k;t,t'|t'',t')=L(k;t,t'|t'',t').\label{eq:31}
\end{equation}
Here
\begin{equation}
L(k;t,t'|t'',t')=M_{ijm}({\bf k})\sum_{{\bf {\bf p+{\bf q=}}k}}\langle u_{j}({\bf p},t)\hat{G}_{mn}({\bf {\bf q}};t,t')\hat{G}_{in}({\bf -{\bf k}};t'',t')\rangle\label{eq:32}
\end{equation}
Here we should mention that Eijnden \cite{Eijnden97} analyzed evolution
equations for $G_{inab}^{G}({\bf k};t,t'|-{\bf k};t'',t')$, to study
joint-statistics of two-particle embedded in a random flow, within
the framework of DIA and his proposed modified-DIA theory.

Now the goal is to obtain VDIA solutions for $S(k;t,t')$ and $L(k;t,t'|t'',t')$. In
the present VDIA theory, usual recipe of DIA {\em with one exception}
is used to obtain closure equations for two-time and single-time velocity
correlations. As an exception and during last step of renormalization,
we use second order correlations $G^{G}$ and substitute $K(k;t,t')$
instead of $G(k;t,t')$. So we do not need equation for $G(k;t,t')$ in VDIA. In general, the following rule is used in
VDIA theory. 
\begin{quotation}
{\bf Rule 1}: All $G$s should be replaced by $K$s in DIA closure
equations for statistical properties (except for equation for $G$ itself) to obtain corresponding VDIA closure
equations.
\end{quotation}
Here we should mention that $K(k;t,t')$ is not affected by random
Galilean velocity field $v_{i}$ and more information is provided
later in this paper. 
The VDIA equations for $U(k;t,t')$ and $U(k;t,t)$
can be obtained by replacing all $G$s to $K$s in DIA Eqs. (\ref{eq:19}),
(\ref{eq:20}) and (\ref{eq:22}), and written as
\begin{equation}
\left(\frac{\partial}{\partial t}+\nu k^{2}\right)U(k;t,t')=S^{VDIA}(k;t,t'),\label{eq:33}
\end{equation}
\begin{equation}
\left(\frac{\partial}{\partial t}+2\nu k^{2}\right)U(k;t,t)=2S^{VDIA}(k;t,t),\label{eq:34}
\end{equation}
where
\begin{eqnarray}
S^{VDIA}(k;t,t') & = & \pi k\int\int_{\Delta}\mathrm{d}p\,\mathrm{d}q\, pq\Bigl[\int_{0}^{t'}\mathrm{d}sa(k,p,q)K(k;t',s)U(p;t,s)U(q;t,s)\nonumber \\
 &  & -\int_{0}^{t}\mathrm{d}sb(k,p,q)U(k;t',s)K(p;t,s)U(q;t,s)\Bigr],\label{eq:35}
\end{eqnarray}

Now we discuss equation for $K(k;t,t')$. Using usual recipe of DIA, the DIA equation for $G_{inab}^{G}({\bf k};t,t'|-{\bf k};t,t')$
can be obtained from Eqs. (\ref{eq:15}), (\ref{eq:2}) and which is provided in the Appendix. And from which VDIA equation for $G^{G}(k;t,t'|t,t')$ can be obtained
by first simplifying the DIA equation for isotropic turbulence and then substituting $K(k;t,s)$ for $G(k;t,s)$ i.e. by using Rule 1.
The final equation for $K(k;t,t')$ when $t>t'$ can be written a
\begin{eqnarray}
\left(\frac{\partial}{\partial t}+2\nu k^{2}\right)\left[K(k;t,t')\right]^{2} & = & 2L^{VDIA}(k;t,t'|t,t')\label{eq:36}
\end{eqnarray}
where
\begin{eqnarray}
L^{VDIA}(k;t,t'|t,t') & = & -\pi\int\int_{\Delta}\mathrm{d}p\mathrm{d}qkpqb(k,p,q)\int_{t'}^{t}\mathrm{d}sK(p;t,s)U(q;t,s)G^{G}(k;t,t'|s,t')\nonumber \\
 & + & 2\pi\int\int_{\Delta}\mathrm{d}p\,\mathrm{d}q\, kpqa(k,p,q)\Bigl[\nonumber \\
 &  & \int_{t'}^{t}\mathrm{d}sK(k;t,s)U(q;t,s)G^{G}(p;t,t'|s,t')\Bigr],\label{eq:37}
\end{eqnarray}
It should be noted that unknown correlation functions of the type
$G^{G}(r;t,t'|t'',t')$, i.e. $G^{G}(k;t,t'|s,t')$ and $G^{G}(p;t,t'|s,t')$
appear in Eq. (\ref{eq:37}). We should mention that Eq. (\ref{eq:37})
is valid for $t>t'$ and $s>t'$. For $s=t',$ $G^{G}(k;t,t'|s,t')$
becomes equal to $G(k;t,t')$ and thus should be replaced by $K(k;t,t')$
in Eq. (\ref{eq:37}) according to Rule 1. Similarly, when $s=t',$ $G^{G}(p;t,t'|s,t')$
should be replaced by $K(p;t,t')$. The VDIA equation for $G^{G}(k;t,t'|t'',t')$
which appears in Eq. (\ref{eq:37}) can be obtained from DIA equation
for $G_{inab}^{G}({\bf k};t,t'|-{\bf k};t'',t')$. The derived DIA
equation for $G_{inab}^{G}({\bf k};t,t'|-{\bf k};t'',t')$ is provided
in the Appendix. From which, the VDIA equation for $G^{G}(k;t,t'|t'',t')$
when $t>t',\, t''>t'$ can be obtained by implementing Rule 1. The
obtained equation can be written as
\begin{eqnarray}
\left(\frac{\partial}{\partial t}+\nu k^{2}\right)G^{G}(k;t,t'|t'',t') & = & L^{VDIA}(k;t,t'|t'',t')
\end{eqnarray}
where
\begin{eqnarray}
L^{VDIA}(k;t,t'|t'',t') & = & -\pi\int\int_{\Delta}\mathrm{d}p\mathrm{d}qkpqb(k,p,q)\int_{t'}^{t}\mathrm{d}sK(p;t,s)U(q;t,s)G^{G}(k;s,t'|t'',t')\nonumber \\
 & + & 2\pi\int\int_{\Delta}\mathrm{d}p\mathrm{d}qkpqa(k,p,q)\Bigl[\nonumber \\
 &  & \int_{t'}^{t''}\mathrm{d}sK(k;t'',s)U(q;t,s)G^{G}(p;t,t'|s,t')\Bigr].\label{eq:39}
\end{eqnarray}
We remind that for $s=t',$ $G^{G}(k;s,t'|t'',t')$ and $G^{G}(p;t,t'|s,t')$
should be replaced by $K(k;t'',t')$ and $K(p;t,t')$, respectively
in Eq. (\ref{eq:39}). These closed set of equations (\ref{eq:33})-(\ref{eq:39})
form closure solution for homogeneous, isotropic turbulence as provided
by VDIA.

The first term on the rhs of Eq. (\ref{eq:37}) and (\ref{eq:39})
is analogous to the term on the rhs of DIA Eq. (\ref{eq:21}). The
second term on the rhs of Eq. (\ref{eq:37}) and (\ref{eq:39}) is
analogous to the first term on the rhs of DIA Eq. (\ref{eq:22}).
This additional second term on the rhs of Eqs. (\ref{eq:37}) and
(\ref{eq:39}) cancels out the divergence of the first term in these
equations. We must mention that the VDIA equations are not compatible
with RGT. In the next section, we suggest modifications to these equations
so as to make them compatible with RGT. Before that, we now
briefly present the approximate analysis of Eqs.(\ref{eq:36}) and (\ref{eq:37})
justifying the absence of divergence. 

\subsection{Approximate Analysis of Equation for $K(k;t,t')$}

Consider stationary homogeneous, isotropic turbulence which is governed
by NS equation with added forcing term. In case of stationary turbulence,
the Eqs. (\ref{eq:36}), (\ref{eq:37}) for $K(k;t,t')$ are still
valid. Also $U(k;t,t)$ and energy spectrum $E(k)$ are independent
of time and are related by
\begin{eqnarray}
\frac{1}{2}U(k;t,t') & = & \frac{1}{4\pi k^{2}}E(k)r(k;t,t'),\label{eq:a77-1}
\end{eqnarray}
where $r(k;t,t')=r(k;t-t')$ is the time correlation function of the
Fourier mode and $r(k;0)=1$ \cite{Kraichnan59,Kraichnan64d}. Following
Kraichnan's approximate analysis for $G$ equation \cite{Kraichnan64d,Leslie73},
we integrate Eqs. (\ref{eq:36}) and (\ref{eq:37}) from $t-t'=0$
to $t-t'=\infty$ and use approximations
\begin{eqnarray}
K(k;t,t') & = & \exp[-\eta(k)(t-t')],\, t\ge t'\label{eq:a78}
\end{eqnarray}
\begin{eqnarray}
r(k;t,t') & = & \exp[-\zeta(k)(t-t')],\, t\ge t'\label{eq:a79-1}
\end{eqnarray}
\begin{eqnarray}
R^{G}(k;t,t'|t'',t')\cong1.\label{eq:a80-1}
\end{eqnarray}
The resulting equation may be written as
\begin{eqnarray}
\eta(k) & = & \nu k^{2}+\frac{k}{2}\int\int_{\Delta}\mathrm{d}p\mathrm{d}qpq^{-1}E(q)\left[\frac{b(k,p,q)\left\{ \eta(p)-\eta(k)\right\} -\eta(k)b(k,q,p)}{\eta(p)\left\{ \eta(k)+\eta(p)+\zeta(q)\right\} }\right].\label{eq:a44-1}
\end{eqnarray}
For conditions $q\ll k$, $E(q)\propto q^{-5/3}$, $\eta(k)\propto k^{2/3}$,
$\zeta(q)\propto q^{2/3}$, the divergence is not present in the first
integral term containing $b(k,p,q)$ on the rhs of Eq. (\ref{eq:a44-1}) due
to the presence of multiplying factor $\eta(p)-\eta(k)$ and the last
term containing $b(k,q,p)$ does not have any divergence. So the equation
for $K(k;t,t')$ is consistent with Kolmogorov's energy spectrum.
A detailed analysis of VDIA theory for stationary, isotropic turbulence
will be presented in the next paper.

\section{random Galilean Transformation Based Modification to VDIA Equations}

We use Kraichnan's concept of random Galilean invariance \cite{Kraichnan64,Kraichnan65a,Pandya14}
and transform Navier-Stokes (NS) equation. The transformed equation hereafter
is referred to as random Galilean Navier-Stokes (RGNS) equation. Now
we present RGNS equation.

In reference frame $S$, we add a uniform Galilean velocity field
$v_{i}$ to a realization of $u_{i}({\bf k},t)$ and obtain a homogeneous
fluctuating turbulent field $u_{i}({\bf k},t)_{v}$ governed by 
\begin{equation}
\left(\frac{\partial}{\partial t}+\nu k^{2}\right)u_{i}({\bf k},t)_{v}=-\iota k_{j}v_{j}u_{i}({\bf k},t)_{v}+M_{ijm}({\bf k})\sum_{{\bf {\bf p+{\bf q=}}k}}u_{j}({\bf p},t)_{v}u_{m}({\bf q},t)_{v}.\label{eq:40}
\end{equation}
Following Kraichnan \cite{Kraichnan65a}, consider $v_{i}$ to be
statistically distributed over the ensemble of infinite realizations
with $\langle v_{j}\rangle=0$ and statistically independent of $u_{i}({\bf k},t)$.
Also, assume Gaussian distribution for $v_{i}$ for later simplicity
with 
\begin{equation}
v_{0}^{2}=\langle v_{1}^{2}\rangle=\langle v_{2}^{2}\rangle=\langle v_{3}^{2}\rangle.
\end{equation}
The Eq. (\ref{eq:40}) with random Galilean velocity $v_{i}$ is referred
to as RGNS equation. Also, the continuity equation is 
\begin{equation}
k_{j}u_{j}({\bf k},t)_{v}=0.\label{eq:42}
\end{equation}

For homogeneous, isotropic turbulence, exact equations for various statistical
properties used in VDIA and now related to $u_{i}({\bf k},t)_{v}$
field may be obtained from Eqs. (\ref{eq:40}) and (\ref{eq:42}).
They can be written as
\begin{equation}
\left[\frac{\partial}{\partial t}+\nu k^{2}\right]U(k;t,t')_{v}=\left(\frac{D}{2\pi}\right)^{3}\langle-\iota k_{j}v_{j}u_{i}({\bf k},t)_{v}u_{i}(-{\bf k},t')_{v}\rangle+S(k;t,t')_{v},\label{eq:43}
\end{equation}
\begin{equation}
\left[\frac{\partial}{\partial t}+2\nu k^{2}\right]U(k;t,t)_{v}=2S(k;t,t)_{v},\label{eq:44}
\end{equation}
\begin{equation}
\left(\frac{\partial}{\partial t}+2\nu k^{2}\right)\left[K(k;t,t')_{v}\right]^{2}=2L(k;t,t'|t,t')_{v},\label{eq:45}
\end{equation}
\begin{eqnarray}
\left(\frac{\partial}{\partial t}+\nu k^{2}\right)G^{G}(k;t,t'|t'',t')_{v} & = & \frac{1}{2}\langle-\iota k_{j}v_{j}\hat{G}_{in}({\bf k};t,t')_{v}\hat{G}_{in}({\bf -k};t'',t')_{v}\rangle\nonumber \\
 &  & +L(k;t,t'|t'',t')_{v}.\label{eq:46}
\end{eqnarray}
The solutions of Eqs. (\ref{eq:2}) and (\ref{eq:40}) are related
by 
\begin{equation}
u_{i}({\bf k},t)_{v}=\exp\left[-\iota k_{j}v_{j}(t-t_{0})\right]u_{i}({\bf k},t)
\end{equation}
 and from which follow the exact random Galilean transformation (RGT)
rules \cite{Kraichnan64,Kraichnan65a,Pandya14}. The RGT rules for statistical properties
appearing in Eqs. (\ref{eq:43})-(\ref{eq:46}) can be written
as 
\begin{equation}
U(k;t,t')_{v}=\exp\left[-\frac{k^{2}v_{0}^{2}}{2}(t-t')^{2}\right]U(k;t,t'),\label{eq:48}
\end{equation}
\begin{equation}
S(k;t,t')_{v}=\exp\left[-\frac{k^{2}v_{0}^{2}}{2}(t-t')^{2}\right]S(k;t,t'),\label{eq:49}
\end{equation}
%
%
%
%
\begin{equation}
G^{G}(k;t,t'|t'',t')_{v}=\exp\left[-\frac{k^{2}v_{0}^{2}}{2}(t-t'')^{2}\right]G^{G}(k;t,t'|t'',t'),\label{eq:50}
\end{equation}
\begin{equation}
L(k;t,t'|t'',t')_{v}=\exp\left[-\frac{k^{2}v_{0}^{2}}{2}(t-t'')^{2}\right]L(k;t,t'|t'',t'),\label{eq:51}
\end{equation}
From Eq. (\ref{eq:50}) we obtain for $t''=t$, 
\begin{equation}
G^{G}(k;t,t'|t,t')_{v}=G^{G}(k;t,t'|t,t'),\; K(k;t,t')_{v}=K(k;t,t')\label{eq:52}
\end{equation}
 and which suggest that $K(k;t,t')$ is not affected by random Galilean
transformation velocity. Using Eqs. (\ref{eq:29-1}), (\ref{eq:50}),
and (\ref{eq:52}) ,we obtain
\begin{equation}
R^{G}(k;t,t'|t'',t')_{v}=\exp\left[-\frac{k^{2}v_{0}^{2}}{2}(t-t'')^{2}\right]R^{G}(k;t,t'|t'',t').\label{eq:53}
\end{equation}
Equations. (\ref{eq:48}), (\ref{eq:49}) and (\ref{eq:51}) suggests
that random Galilean transformation velocity does not affect $U(k;t,t)$,
$S(k;t,t)$ and $L(k;t,t'|t,t')$, which can be written as 
\begin{equation}
U(k;t,t)_{v}=U(k;t,t),\label{eq:54}
\end{equation}
\begin{equation}
S(k;t,t)_{v}=S(k;t,t),\label{eq:55}
\end{equation}
\begin{equation}
L(k;t,t'|t,t')_{v}=L(k;t,t'|t,t'),\label{eq:56}
\end{equation}

\subsection{Modified equations for $U(k;t,t)$ and $K(k;t,t')$}

Using Eqs. (\ref{eq:48}), (\ref{eq:50}), (\ref{eq:52}), it can
be verified that the VDIA Eqs. (\ref{eq:34})-(\ref{eq:37}) are not
compatible with RGT because
\begin{equation}
S^{VDIA}(k;t,t)_{v}\ne S^{VDIA}(k;t,t),\label{eq:55-1}
\end{equation}
\begin{equation}
L^{VDIA}(k;t,t)_{v}\ne L^{VDIA}(k;t,t),\label{eq:56-1}
\end{equation}
and which violate the rules of RGT as given by Eqs. (\ref{eq:55})
and (\ref{eq:56}). We now suggests modified VDIA equations for $U(k;t,t)$
and $K(k;t,t')$ by introducing adjustable functions $\Gamma_{1}^{U}$,
$\Gamma_{2}^{U}$, $\Gamma_{1}^{G}$, $\Gamma_{2}^{G}$ in the expressions
for the $S^{VDIA}(k;t,t)$ and $L^{VDIA}(k;t,t).$ Then we suggest
different possibilities for these functions which would make the modified
equations invariant under RGT. The modified equations can be written
as
\begin{eqnarray}
\left(\frac{\partial}{\partial t}+2\nu k^{2}\right)U(k;t,t) & = & 2\pi k\int\int_{\Delta}\mathrm{d}p\mathrm{d}qpq\Bigl[\int_{0}^{t}a(k,p,q)K(k;t,s)U(p;t,s)U(q;t,s)\Gamma_{1}^{U}\mathrm{d}s\nonumber \\
 &  & -\int_{0}^{t}b(k,p,q)U(k;t,s)K(p;t,s)U(q;t,s)\Gamma_{2}^{U}\mathrm{d}s\Bigr],\label{eq:43-1}
\end{eqnarray}
and 
\begin{eqnarray}
\left(\frac{\partial}{\partial t}+2\nu k^{2}\right)\left[K(k;t,t')\right]^{2} & = & -2\pi\int\int_{\Delta}\mathrm{d}p\mathrm{d}qkpqb(k,p,q)\Bigl[\nonumber \\
 &  & \int_{t'}^{t}K(p;t,s)U(q;t,s)G^{G}(k;t,t'|s,t')\Gamma_{2}^{G}\mathrm{d}s\Bigr]\nonumber \\
 &  & +4\pi\int\int_{\Delta}\mathrm{d}p\mathrm{d}qkpqa(k,p,q)\Bigl[\nonumber \\
 &  & \int_{t'}^{t}K(k;t,s)U(q;t,s)G^{G}(p;t,t'|s,t')\Gamma_{1}^{G}\mathrm{d}s\Bigr].\label{eq:60}
\end{eqnarray}
A possible choice for functions $\Gamma_{1}^{U}$, $\Gamma_{2}^{U}$,
$\Gamma_{1}^{G}$, $\Gamma_{2}^{G}$ can be 
\begin{equation}
\Gamma_{1}^{U}=\frac{1}{R^{G}\left(p;t,0|s,0\right)R^{G}\left(q;t,0|s,0\right)},\; t>0,\: s>0,\label{eq:61}
\end{equation}
\begin{equation}
\Gamma_{2}^{U}=\frac{1}{R^{G}\left(k;t,0|s,0\right)R^{G}\left(q;t,0|s,0\right)},\; t>0,\: s>0,\label{eq:62}
\end{equation}
\begin{equation}
\Gamma_{1}^{G}=\frac{1}{R^{G}\left(p;t,t'|s,t'\right)R^{G}\left(q;t,t'|s,t'\right)},\; t>t',\: s>t',\label{eq:63}
\end{equation}
\begin{equation}
\Gamma_{2}^{G}=\frac{1}{R^{G}\left(k;t,t'|s,t'\right)R^{G}\left(q;t,t'|s,t'\right)},\; t>t',\: s>t'.\label{eq:64}
\end{equation}
It should be noted that Eqs. (\ref{eq:61}), (\ref{eq:62}) are valid for $s>0$ and Eqs. (\ref{eq:63}), (\ref{eq:64}) are valid for $s>t'$. So for $s=0$ and $s=t'$ which are lower limits of integrals in Eqs. (\ref{eq:43-1}) and (\ref{eq:60}), conditions of RGT are not satisfied. We should mention that these adjustable functions remain valid when the distribution of random Galilean velocity $v_i$ is not Gaussian. In case of non-Gaussian distribution of $v_i$, the factor $\exp\left[-\frac{k^{2}v_{0}^{2}}{2}(t-t')^{2}\right]$ should be replaced by $\langle\exp\left[-\iota k_jv_j(t-t')\right]\rangle$ in various equations of RGT rules.  

When $v_i$ has Gaussian distribution, another possible choice for functions $\Gamma_{1}^{U}$, $\Gamma_{2}^{U}$,
$\Gamma_{1}^{G}$, $\Gamma_{2}^{G}$ can be 
\begin{equation}
\Gamma_{1}^{U}=\frac{1}{R^{G}\left(\sqrt{p^{2}+q^{2}};t,0|s,0\right)},\; t>0,\: s>0
\end{equation}
\begin{equation}
\Gamma_{2}^{U}=\frac{1}{R^{G}\left(\sqrt{k^{2}+q^{2}};t,0|s,0\right)},\; t>0,\: s>0,
\end{equation}
\begin{equation}
\Gamma_{1}^{G}=\frac{1}{R^{G}\left(\sqrt{p^{2}+q^{2}};t,t'|s,t'\right)},\; t>t',\: s>t',
\end{equation}
\begin{equation}
\Gamma_{2}^{G}=\frac{1}{R^{G}\left(\sqrt{k^{2}+q^{2}};t,t'|s,t'\right)},\; t>t',\: s>t'.
\end{equation}
We remind that these functions are valid for $s>0$ and $s>t'$ and conditions of RGT are not satisfied in Eqs. (\ref{eq:43-1}) and (\ref{eq:60}) when $s=0$ and $s=t'$, respectively.

\subsection{Modified equations for $U(k;t,t')$ and $G^{G}(k;t,t'|t'',t')$}

It is easy to verify that the VDIA equations for $U(k;t,t'$) and
$G^{G}(k;t,t'|t'',t')$ are not compatible with RGT as
\begin{equation}
S^{VDIA}(k;t,t')_{v}\ne\exp\left[-\frac{k^{2}v_{0}^{2}}{2}(t-t')^{2}\right]S^{VDIA}(k;t,t'),
\end{equation}
\begin{equation}
L^{VDIA}(k;t,t'|t'',t')_{v}\ne\exp\left[-\frac{k^{2}v_{0}^{2}}{2}(t-t'')^{2}\right]L^{VDIA}(k;t,t'|t'',t'),\label{eq:70}
\end{equation}
and which violate the rules of RGT as given by Eqs. (\ref{eq:49})
and (\ref{eq:51}). Consider following modified equations:
\begin{eqnarray}
\left(\frac{\partial}{\partial t}+\nu k^{2}\right)U(k;t,t') & = & \pi k\int\int_{\Delta}\mathrm{d}p\mathrm{d}qpq\Bigl[\int_{0}^{t'}a(k,p,q)K(k;t',s)U(p;t,s)U(q;t,s)\Lambda_{1}^{U}\mathrm{d}s\nonumber \\
 &  & -\int_{0}^{t}b(k,p,q)U(k;t',s)K(p;t,s)U(q;t,s)\Lambda_{2}^{U}\mathrm{d}s\Bigr],\label{eq:71}
\end{eqnarray}
\begin{eqnarray}
\left(\frac{\partial}{\partial t}+\nu k^{2}\right)G^{G}(k;t,t'|t'',t') & = & -\pi\int\int_{\Delta}\mathrm{d}p\mathrm{d}qkpqb(k,p,q)\Bigl[\nonumber \\
 &  & \int_{t'}^{t}K(p;t,s)U(q;t,s)G^{G}(k;s,t'|t'',t')\Lambda_{2}^{G}\mathrm{d}s\Bigr]\nonumber \\
 &  & +2\pi\int\int_{\Delta}\mathrm{d}p\mathrm{d}qkpqa(k,p,q)\Bigl[\nonumber \\
 &  & \int_{t'}^{t''}K(k;t'',s)U(q;t,s)G^{G}(p;t,t'|s,t')\Lambda_{1}^{G}\mathrm{d}s\Bigr].\label{eq:72}
\end{eqnarray}
A possible choice for functions $\Lambda_{1}^{U}$, $\Lambda_{2}^{U}$,
$\Lambda_{1}^{G}$, $\Lambda_{2}^{G}$ for satisfying RGT conditions
is
\begin{equation}
\Lambda_{1}^{U}=\frac{R^{G}(k;t,0|t',0)}{R^{G}\left(p;t,0|s,0\right)R^{G}\left(q;t,0|s,0\right)},\; t>0,\: s>0,\, t'>0,\label{eq:73}
\end{equation}
\begin{equation}
\Lambda_{2}^{U}=\frac{R^{G}(k;t,0|t',0)}{R^{G}\left(k;t',0|s,0\right)R^{G}\left(q;t,0|s,0\right)},\; t>0,\: s>0,\, t'>0,\label{eq:74}
\end{equation}
\begin{equation}
\Lambda_{1}^{G}=\frac{R^{G}(k;t,s|t'',s)}{R^{G}\left(p;t,t'|s,t'\right)R^{G}\left(q;t,t'|s,t'\right)},\; t>t',\: s>t',\: s<t'',\:s<t\label{eq:75}
\end{equation}
\begin{equation}
\Lambda_{2}^{G}=\frac{R^{G}(k;t,s|t'',s)}{R^{G}\left(k;t'',t'|s,t'\right)R^{G}\left(q;t,t'|s,t'\right)},\; t>t',\: s>t',\:s<t'',\:s<t.\label{eq:76}
\end{equation}
It should be noted that for $t'=t$, Eqs. (\ref{eq:73}), (\ref{eq:74})
for $\Lambda_{1}^{U}$, $\Lambda_{2}^{U}$ become identical to Eqs.
(\ref{eq:61}), (\ref{eq:62}) for $\Gamma_{1}^{U}$, $\Gamma_{2}^{U}$,
respectively. Also, for $t''=t$, Eqs. (\ref{eq:75}), (\ref{eq:76})
for $\Lambda_{1}^{G}$, $\Lambda_{2}^{G}$ become identical to Eqs.
(\ref{eq:63}), (\ref{eq:64}) for $\Gamma_{1}^{G}$, $\Gamma_{2}^{G}$,
respectively. When $s=0$, RGT conditions are not satisfied in Eq. (\ref{eq:71}). And for $s=t',s=t'',s=t$, conditions of RGT are not satisfied in Eq. (\ref{eq:72}). We should mentoin that these adjustable functions remain valid in case of non-Gaussian distribution for $v_i$. 

For future reference, hereafter we refer to these RGT
compatible (with a few restrictions as discussed above) modified VDIA equations as Invariant-VDIA (I-VDIA) equations.

\section{Concluding Remarks}

In this paper, we have solved closure problem of turbulence by suggesting a new Eulerian theory, namely VDIA, within the framwork of Kraichnan's DIA theory. The VDIA theory is cosistent with Kolmogorov's energy spectrum. The VDIA equation for $K(k;t,t')$  does not have divergence and is compatible with the scaling $-5/3$ of Kolmogorov's energy spectrum. Further, a modified version of VDIA, namely, I-VDIA is suggested in which concepts of random Galilean transformation are implemented. The I-VDIA equations are compatible with rules of random Galilean transformation, except at time values which are limits of integral terms in these equations. We should mention that VDIA and I-VDIA do preserve the conservation of energy among any triad interaction of Fourier modes \cite{Kraichnan59}.  The analysis of VDIA theory in case of stationary, isotropic turbulence will be presented in the next paper. Also, the numerical assessment of VDIA and I-VDIA against numerical solution of NS equation will be considered in future work.

\section*{Appendix}

The DIA equation for $G_{inab}^{G}({\bf k};t,t'|-{\bf k};t,t')$ for $t>t'$ can
be written as
\begin{eqnarray}
\left(\frac{\partial}{\partial t}+2\nu k^{2}\right)G_{inab}^{G}({\bf k};t,t'|-{\bf k};t,t') & = & L_{inab}({\bf k},-{\bf k};t,t')\label{eq:77-2}
\end{eqnarray}
where
\begin{eqnarray}
L_{inab}({\bf k},-{\bf k};t,t') & = & 4M_{ijm}(\mathbf{k})\sum_{{\bf {\bf p+{\bf q=}}k}}\Bigl[M_{sdc}(-\mathbf{k})\nonumber \\
 &  & \int_{t'}^{t}G_{as}(-\mathbf{k};t,t'')Q_{jd}(\mathbf{p},-\mathbf{p};t,t'')G_{mncb}^{G}({\bf q};t,t'|-{\bf q};t'',t')\mathrm{d}t''\nonumber \\
 & + & M_{sdc}({\bf q})\int_{t'}^{t}G_{ms}(\mathbf{q};t,t'')Q_{jd}(\mathbf{p},-\mathbf{p};t,t'')G_{cnab}^{G}({\bf k};t'',t'|-{\bf k};t,t')\mathrm{d}t''\Bigr]\nonumber \\
 & + & 4M_{ajm}(-\mathbf{k})\sum_{{\bf {\bf p+{\bf q=}-}k}}\Bigl[M_{sdc}(\mathbf{k})\nonumber \\
 &  & \int_{t'}^{t}G_{is}(\mathbf{k};t,t'')Q_{jd}(\mathbf{p},-\mathbf{p};t,t'')G_{mbcn}^{G}({\bf q};t,t'|-{\bf q};t'',t')\mathrm{d}t''\nonumber \\
 & + & M_{sdc}({\bf q})\int_{t'}^{t}G_{ms}(\mathbf{q};t,t'')Q_{jd}(\mathbf{p},-\mathbf{p};t,t'')G_{incb}^{G}({\bf k};t,t'|-{\bf k};t'',t')\mathrm{d}t''\Bigr].\label{eq:77}
\end{eqnarray}

The DIA equation for $G_{inab}^{G}({\bf k};t,t'|-{\bf k};t'',t')$ for $t>t',t''>t'$
can be written as
\begin{eqnarray}
\left(\frac{\partial}{\partial t}+2\nu k^{2}\right)G_{inab}^{G}({\bf k};t,t'|-{\bf k};t'',t') & = & J_{inab}({\bf k},-{\bf k};t,t',t'')\label{eq:79}
\end{eqnarray}
where
\begin{eqnarray}
J_{inab}({\bf k},-{\bf k};t,t',t'') & = & 4M_{ijm}(\mathbf{k})\sum_{{\bf {\bf p+{\bf q=}}k}}\Bigl[M_{sdc}(-\mathbf{k})\nonumber \\
 &  & \int_{t'}^{t''}G_{as}(-\mathbf{k};t'',s)Q_{jd}(\mathbf{p},-\mathbf{p};t,s)G_{mncb}^{G}({\bf q};t,t'|-{\bf q};s,t')\mathrm{d}s\Bigr]\nonumber \\
 & + & 4M_{ijm}(\mathbf{k})\sum_{{\bf {\bf p+{\bf q=}}k}}\Bigl[M_{sdc}({\bf q})\nonumber \\
 &  & \int_{t'}^{t}G_{ms}(\mathbf{q};t,s)Q_{jd}(\mathbf{p},-\mathbf{p};t,s)G_{abcn}^{G}(-{\bf k};t'',t'|{\bf k};s,t')\mathrm{d}s\Bigr].\label{eq:80}
\end{eqnarray}

\makeatletter \providecommand{\@ifxundefined}[1]{%
 \@ifx{#1\undefined}
}\providecommand{\@ifnum}[1]{%
 \ifnum #1\expandafter \@firstoftwo
 \else \expandafter \@secondoftwo
 \fi
}\providecommand{\@ifx}[1]{%
 \ifx #1\expandafter \@firstoftwo
 \else \expandafter \@secondoftwo
 \fi
}\providecommand{\natexlab}[1]{#1}\providecommand{\enquote}[1]{``#1''}\providecommand{\bibnamefont}[1]{#1}\providecommand{\bibfnamefont}[1]{#1}\providecommand{\citenamefont}[1]{#1}\providecommand{\href@noop}[0]{\@secondoftwo}\providecommand{\href}[0]{\begingroup \@sanitize@url \@href}\providecommand{\@href}[1]{\@@startlink{#1}\@@href}\providecommand{\@@href}[1]{\endgroup#1\@@endlink}\providecommand{\@sanitize@url}[0]{\catcode `\\12\catcode `\$12\catcode
  `\&12\catcode `\#12\catcode `\^12\catcode `\_12\catcode `\%12\relax}\providecommand{\@@startlink}[1]{}\providecommand{\@@endlink}[0]{}\providecommand{\url}[0]{\begingroup\@sanitize@url \@url }\providecommand{\@url}[1]{\endgroup\@href {#1}{\urlprefix }}\providecommand{\urlprefix}[0]{URL }\providecommand{\Eprint}[0]{\href }\providecommand{\doibase}[0]{http://dx.doi.org/}\providecommand{\selectlanguage}[0]{\@gobble}\providecommand{\bibinfo}[0]{\@secondoftwo}\providecommand{\bibfield}[0]{\@secondoftwo}\providecommand{\translation}[1]{[#1]}\providecommand{\BibitemOpen}[0]{}\providecommand{\bibitemStop}[0]{}\providecommand{\bibitemNoStop}[0]{.\EOS\space}\providecommand{\EOS}[0]{\spacefactor3000\relax}\providecommand{\BibitemShut}[1]{\csname bibitem#1\endcsname}\let\auto@bib@innerbib\@empty

\bibliographystyle{prsty}

\begin{thebibliography}{10}

\bibitem{Kraichnan58}
R.~H. Kraichnan, Phys. Rev. {\bf 109},  1407  (1958).

\bibitem{Kraichnan59}
R.~H. Kraichnan, J. Fluid Mech. {\bf 5},  497  (1959).

\bibitem{KH78}
R.~H. Kraichnan and J.~R. Herring, J. Fluid Mech. {\bf 88},  355  (1978).

\bibitem{Wyld61}
H.~W. {Wyld Jr.}, Ann. Phys. {\bf 14},  143  (1961).

\bibitem{Edwards64}
S.~F. Edwards, J. Fluid Mech. {\bf 18},  239  (1964).

\bibitem{Kraichnan65a}
R.~H. Kraichnan, Phys. Fluids {\bf 8},  575  (1965).

\bibitem{Lee65}
L.~L. Lee, Ann. Phys. {\bf 32},  292  (1965).

\bibitem{Herring65}
J.~R. Herring, Phys. Fluids {\bf 8},  2219  (1965).

\bibitem{Herring66}
J.~R. Herring, Phys. Fluids {\bf 9},  2106  (1966).

\bibitem{EM69}
S.~F. Edwards and W.~D. McComb, J. Phys. A {\bf 2},  157  (1969).

\bibitem{Phythian69}
R. Phythian, J. Phys. A {\bf 2},  181  (1969).

\bibitem{BS70}
R. Balescu and A. Senatorski, Ann. Phys. {\bf 58},  587  (1970).

\bibitem{Orszag70}
S.~A. Orszag, J. Fluid Mech. {\bf 41},  363  (1970).

\bibitem{Nakano72}
T. Nakano, Ann. Phys. {\bf 73},  326  (1972).

\bibitem{Nakano88}
T. Nakano, Phys. Fluids. {\bf 31},  1420  (1988).

\bibitem{MSR73}
P.~C. Martin, E.~D. Siggia, and H.~A. Rose, Phys. Rev. A {\bf 8},  423  (1973).

\bibitem{FLB74}
U. Frisch, M. Lesieur, and A. Brissaud, J. Fluid Mech. {\bf 65},  145  (1974).

\bibitem{FNS77}
D. Forster, D.~R. Nelson, and M.~J. Stephen, Phys. Rev. A {\bf 16},  732
  (1977).

\bibitem{McComb78}
W.~D. McComb, J. Phys. A {\bf 11},  613  (1978).

\bibitem{DM79}
C. DeDominicis and P.~C. Martin, Phys. Rev. A {\bf 19},  419  (1979).

\bibitem{HL79}
H. Horner and R. Lipowsky, Z. Physik B {\bf 33},  223  (1979).

\bibitem{Kaneda81}
Y. Kaneda, J. Fluid Mech. {\bf 107},  131  (1981).

\bibitem{AVP83}
L.~T. Adzhemyan, A.~N. Vasiliev, and Y.~M. Pis{'}mak, Theor. Math. Phys. {\bf
  57},  1131  (1983).

\bibitem{Qian83}
J. Qian, Phys. Fluids {\bf 26},  2098  (1983).

\bibitem{Lvov91}
V.~S. L'vov, Phys. Rep. {\bf 207},  1  (1991).

\bibitem{BKO93}
J.~C. Bowman, J.~A. Krommes, and M. Ottaviani, Phys. Fluids B {\bf 5},  3558
  (1993).

\bibitem{MW95}
C. Mou and P.~B. Weichman, Phys. Rev. E {\bf 52},  3738  (1995).

\bibitem{LP95}
V.~S. L'vov and I. Procaccia, Phys. Rev. E. {\bf 52},  3840  (1995).

\bibitem{LP95b}
V.~S. L'vov and I. Procaccia, Phys. Rev. E. {\bf 52},  3858  (1995).

\bibitem{Eijnden97}
E.~V. Eijnden, Ph.D Thesis\, Universite Libre De Bruxelles  (1997).

\bibitem{KG97}
S. Kida and S. Goto, J. Fluid Mech. {\bf 345},  307  (1997).

\bibitem{Pandya04a}
R.~V.~R. Pandya, Phys. Rev. E {\bf 70},  066307  (2004).

\bibitem{Pandya14}
R.~V.~R. Pandya, arXiv:1307.2324  1  (2014).

\bibitem{Leslie73}
D.~C. Leslie, {\em Developments in the Theory of Turbulence} (Clarendon Press,
  Oxford, 1973).

\bibitem{McComb90}
W.~D. McComb, {\em The Physics of Fluid Turbulence} (Oxford University Press,
  New York, NY, 1990).

\bibitem{McComb95}
W.~D. McComb, Rep. Prog. Phys. {\bf 58},  1117  (1995).

\bibitem{Lesieur97}
M. Lesieur, {\em Turbulence in Fluids}, 3rd ed. (Kluwer, Dordrecht, 1997).

\bibitem{Krommes97}
J.~A. Krommes, Physics Reports {\bf 283},  5  (1997).

\bibitem{Yoshizawa98}
A. Yoshizawa, {\em Hydrodynamic and Magnetohydrodynamic Turbulent Flows}
  (Springer Science+Business Media, B. V., Dordrecht, 1998).

\bibitem{AAV99}
L.~T. Adzhemyan, N.~V. Antonov, and A.~N. Vasiliev, {\em Field Theoretic
  Renormalization Group in Fully Developed Turbulence} (Gordon and Breach, The
  Netherlands, 1999).

\bibitem{Kraichnan61}
R.~H. Kraichnan, J. Math. Phys. {\bf 2},  124  (1961).

\bibitem{Kraichnan70a}
R.~H. Kraichnan, J. Fluid Mech. {\bf 41},  189  (1970).

\bibitem{Kraichnan64b}
R.~H. Kraichnan, Phys. Fluids {\bf 7},  1048  (1964).

\bibitem{Kraichnan64c}
R.~H. Kraichnan, Phys. Fluids {\bf 7},  1169  (1964).

\bibitem{Frederiksen99}
J.~S. Frederiksen, J. Atmos. Sci. {\bf 56},  1481  (1999).

\bibitem{OF2004}
T.~J. O'Kane and J.~S. Frederiksen, J. Fluid Mech. {\bf 504},  133  (2004).

\bibitem{OF2010}
T.~J. O'Kane and J.~S. Frederiksen, Frontiers of Fundamental and Computational
  Physics {-} Proc. 10th Int. Symposium  191  (2010).

\bibitem{Kolmogorov41}
A.~N. Kolmogorov, Dokl. Akad. Nauk SSSR {\bf 30},  301  (1941).

\bibitem{Batchelor59}
G.~K. Batchelor, {\em The Theory of Homogeneous Turbulence} (Cambridge
  University Press, Cambridge, UK, 1959).

\bibitem{Frisch95}
U. Frisch, {\em Turbulence} (Cambridge University Press, New York, NY, 1995).

\bibitem{Kraichnan64}
R.~H. Kraichnan, Phys. Fluids {\bf 7},  1723  (1964).

\bibitem{HK79}
J.~R. Herring and R.~H. Kraichnan, J. Fluid Mech.  581  .

\bibitem{Reeks91}
M.~W. Reeks, Phys. Fluids {\bf 3},  446  (1991).

\bibitem{Reeks92}
M.~W. Reeks, Phys. Fluids {\bf 4},  1290  (1992).

\bibitem{BL87}
V.~I. Belinicher and V.~S. L'vov, Sov. Phys. JETP {\bf 66},  303  (1988).

\bibitem{Kraichnan64a}
R.~H. Kraichnan, Phys. Fluids {\bf 7},  1030  (1964).

\bibitem{Kraichnan64d}
R.~H. Kraichnan, Phys. Fluids {\bf 7},  1163  (1964).

\end{thebibliography}

\end{document}